\documentclass[aps,prl,twocolumn,groupedaddress,amssymb,showpacs]{revtex4}
\usepackage{graphicx}
\usepackage{bm}

\begin{document}

\preprint{CMU-HEP-02-04}

\title{Radion Induced Brane Preheating}

\author{Hael Collins}
\email{hael@cmuhep2.phys.cmu.edu}
\author{R.~Holman}
\email{holman@cmuhep2.phys.cmu.edu}
\author{Matthew R.~Martin}
\email{mmartin@cmu.edu}
\affiliation{Department of Physics, Carnegie Mellon University, Pittsburgh, PA
15213}

\date{\today}

\begin{abstract}
When the interbrane separation in compact Randall-Sundrum models is stabilized
via the Goldberger-Wise mechanism, a potential is generated for the four
dimensional field that encodes this geometric information, the so-called {\it
radion\/}.  Due to its origin as a part of the full five dimensional metric,
the radion couples directly to particles on both branes. We exhibit the
exponential growth in the number of brane particles due to parametric
amplification from radion oscillations and discuss some of the consequences of
this process for brane cosmology.

\end{abstract}


\maketitle

Compact Randall-Sundrum models (RS1) \cite{rsa} provide an extremely
interesting relation between hierarchies in four dimensional physics and the
warping of the five dimensional bulk geometry in which our universe is
embedded.  These models consist of two 3-branes placed at the fixed points of
an $S^1/\mathbb{Z}_2$ orbifold, which serve as the boundaries of a slice of a
five dimensional anti-de Sitter space-time (AdS$_5$).  In order to balance the
bulk cosmological constant, the branes must have equal and opposite tensions,
which are in turn related to the value of the bulk cosmological constant.

These models necessarily contain a geometrical mode that encodes the physical
separation between the branes, the so-called {\it radion\/}. We can write the
full 5-d line element in a way that exhibits it explicitly \cite{cgr},
\begin{eqnarray}
ds^2 &=& e^{-2k\left[ y + f(x)e^{2ky}\right]}
\hat g_{\mu\nu}(x)\, dx^\mu dx^\nu \nonumber \\
& & + \left( 1 + 2k f(x)e^{2k y} \right)^2\, dy^2.
\label{cgrmetric}
\end{eqnarray}
The radion is given by the function $f(x)$ and the metric is written in such a
way as to properly incorporate the symmetries of the geometric setup.  In
particular, this form guarantees that the radion will not mix with massless
degrees of freedom such as the $3+1$ gravitons \cite{cgr}.

From the linearized Einstein equations
\begin{equation}
\hat R_{\mu\nu} - {\textstyle{1\over 2}} \hat g_{\mu\nu} \hat R = 0
\qquad\quad \hat\nabla^2 f = 0,
\label{4dEIN}
\end{equation}
where $\hat R_{\mu\nu}$ and $\hat\nabla_\mu$ are the Ricci tensor and
covariant derivative associated with the metric $\hat g_{\mu\nu}(x)$, we can
see that the radion behaves as a massless 4-d field.  This is in accord with
its status as the Goldstone boson corresponding to the fact that the action is
invariant under the operation of placing the second brane at {\it any\/}
distance from the first.  Furthermore, it should be noted that the radion has
{\it no\/} Kaluza-Klein tower associated with it \cite{cgr,noKK}.

The Goldstone nature of the radion precludes the system from picking a value
for the interbrane separation.  Physically, this separation controls the
hierarchy between particle physics scales and that of gravity, so that in
order to make the RS1 scenario work, some means of fixing the expectation
value of the radion must be given.  This is the essence of the Goldberger-Wise
(GW) mechanism \cite{gw} which introduces a bulk scalar with a non-trivial
profile in the bulk and potentials on each of the branes.  For simplicity, the
self-couplings of the GW scalar were chosen large enough so as to ``stiffen''
the potentials to the point that the value of the GW scalar on each brane is
completely fixed.  Integrating the action of the GW scalar over the bulk
coordinate gives rise to a potential for the radion, whose expectation value
can then be fixed in terms of the parameters of the GW action to take the
correct value so as to enforce the required hierarchy.

Our interest in the radion was piqued by the fact that since it {\it is\/} a
mode contained in the full metric, it should couple to all brane fields (or at
least those whose actions break Weyl invariance on the brane) in a universal
manner.  From a cosmological perspective, we should expect that the radion
will not settle to its minimum instantaneously, but that it will oscillate for
some time about the minimum of the potential generated by the GW mechanism. 
As with the inflaton at the end of the inflationary era \cite{inflation},
these oscillations can induce the \emph{explosive} production of brane
particles via parametric amplification, i.e.~{\it preheating\/}
\cite{parametricinflation}. The energy density and equation of state of these
particles will influence the cosmological evolution of the relevant brane. We
calculate this particle production below for the case of a brane scalar
coupled to the radion.

We expect that when the back reaction on the radion due to this particle
production is taken into account, the radion evolution will become dissipative
and its expectation value will settle to its minimum.  However, this is beyond
the scope of this paper and will be treated, together with other related
topics, in \cite{later}.

In order to calculate brane preheating effects due to radion oscillations, we
first need the radion action.  The kinetic terms come from the five
dimensional Einstein action, supplemented by the Gibbons-Hawking terms on the
boundary branes which contribute to the effective potential for the radion.
The GW action is that of a free scalar field of mass $m$ with potentials to
fix the value of the field to be $v_0M_5^{3/2}$, and $v_1M_5^{3/2}$ on the UV
and the IR branes, respectively, where $M_5$ is the five dimensional Planck
mass. The calculation of the radion effective action obtained by integrating
out the extra dimension is relatively straightforward and will be given in
detail in \cite{later}; the resulting effective Lagrangian is
\begin{eqnarray}
{\cal L}_{\rm radion} &=& - \left( 1 - {F\over\Lambda_W} +
{1\over 2} {F^2\over\Lambda^2_W} \right)
\partial^\mu F\partial_\mu F \nonumber \\
&&- {1\over 2} m_F^2 F^2 + {4\over 3} g_F \Lambda_W F^3 - g_F F^4 + \cdots.
\label{radionLag}
\end{eqnarray}
We have defined the following quantities in Eq.~(\ref{radionLag}):
\begin{eqnarray}
F(x) &=& \sqrt{12kM_5^3} e^{k\Delta y} f(x) \nonumber \\
\Lambda_W &\equiv& e^{-k\Delta y} \sqrt{{3M_5^3\over k}}
\sim {\cal O} ({\rm TeV}) \nonumber \\
m_F &=& {v_1(1-\eta)^{1/4}\over 3} \left({m\over M_5}\right)^{3/2} \Lambda_W
\nonumber \\
g_F &\equiv& {1\over 2} {m_F^2\over\Lambda_W^2} ,
\label{parameters}
\end{eqnarray}
where $\Delta y$ is the location of the IR brane relative to the UV brane at
$y=0$, $\Lambda_W$ is the cutoff for the effective theory of the radion such
that at energies above $\Lambda_W$ the radion dynamics become strong and
$\eta$ $(<1)$ parameterizes the change in the tension of the IR brane in order
to account for the effects of the GW scalar.  Ref.~\cite{cgk} also found an
${\mathcal O}({\rm TeV})$ mass for the radion based on a calculation in the
full theory rather than the effective action approach which yielded
Eq.~(\ref{radionLag}).

We envisage a scenario in which at the early stages of an RS universe, when
the typical temperature of the bulk is still above $e^{-k\Delta y}M_5$, the
space-time is generally unstable to the formation of a black hole horizon in
the bulk whose Hawking temperature is the same as that of the bulk.  The model
then resembles the Randall-Sundrum model \cite{rsb} with a single UV brane and
a horizon.  As the universe cools below the energy scale associated with the
IR brane, the free energy of the model with two branes becomes lower
\cite{cnr}.  After its formation it will generally not be at its equilibrium
position but will oscillate about it with a natural amplitude of the order of
$\Lambda_W$, thus driving the brane preheating process.

Thus consider a scalar field $\Phi(x^{\mu})$ confined to the IR brane at
$y=\Delta y$.  Its action involves the radion mode through the induced metric. 
In the small amplitude regime for $\Phi$, we can write the scalar action as
\begin{equation}
S_\Phi = \int_{y=\Delta y} \!\!\!\!\!\!\! d^4x\, \sqrt{-h}\,
\left[ - {\textstyle{1\over 2}} h^{ab} \partial_a\tilde\Phi
\partial_b\tilde\Phi - {\textstyle{1\over 2}} \tilde m_\Phi^2
\tilde\Phi^2 + \cdots \right] , \label{Phiaction}
\end{equation}
where $h_{ab}$ is the induced metric on the IR brane.  Using
Eq.~(\ref{cgrmetric}), we can rewrite this as
\begin{eqnarray}
S_\Phi &=& \int d^4x\, \sqrt{-\hat g}\, \Bigl[
- {\textstyle{1\over 2}} e^{-F/\Lambda_W} \hat g^{\mu\nu}
\partial_\mu\Phi\partial_\nu\Phi \nonumber \\
& &\qquad\qquad\qquad - {\textstyle{1\over 2}} e^{-2F/\Lambda_W}
m_\Phi^2\Phi^2 + \cdots \Bigr] , \quad
\label{branescalar}
\end{eqnarray}
where we have performed the rescalings $\Phi = e^{-k\Delta y}\tilde\Phi$ and
$m_\Phi = e^{-k\Delta y}\tilde m_\Phi$ so as to obtain a canonically
normalized kinetic term when the radion is at its equilibrium value $F=0$. The
couplings of the radion to Standard Model fields were also considered in
\cite{cgk} where precision electroweak observables were studied. In contrast,
we are more concerned with dynamical aspects of these couplings of the radion
to brane fields.

The oscillating radion acts as a time dependent background in which $\Phi$
evolves, i.e.~we neglect radion quantum fluctuations as well as the effects of
the back-reaction due to the produced particles or other fields.  This should
be a reasonable assumption when the back-reaction on the radion from other
fields is small and the higher order terms in the effective radion Lagrangian
Eq.~(\ref{radionLag}) can be neglected.  The time dependence of the radion
background and its couplings to $\Phi$ imply that energy can be drained off
the radion to produce $\Phi$ quanta.  In the absence of back-reaction this
process proceeds indefinitely.  A more realistic expectation is that the
back-reaction will damp out the motion of the radion \cite{later}.  We take
the radion background to be of the form
\begin{equation}
F(t) = F_0 \Lambda_W \cos(m_F t),
\label{oscillating}
\end{equation}
where $F_0$ is dimensionless.

To compute the number of particles produced, we expand $\Phi$ in terms of
creation and annihilation operators, $a_{\vec k}^\dagger$ and $a_{\vec k}$ for
a mode with momentum $\vec k$,
\begin{equation}
\Phi(t,\vec x) = \int {d^3k\over (2\pi)^{3/2}}\, \left[ a_{\vec k}
\Phi_k(t) e^{-i\vec k\cdot\vec x} + a_{\vec k}^\dagger \Phi_k^*(t)
e^{i\vec k\cdot\vec x} \right] .
\label{Phimodes}
\end{equation}
The mode functions are then determined by the equation of motion for the
scalar field from Eq.~(\ref{branescalar}),
\begin{equation}
{d^2\Phi_{\vec k}\over dt^2} + F_0 m_F \sin(m_F t) {d\Phi_{\vec
k}\over dt} + {\omega}^2_k (t) \Phi_{\vec k} = 0,
\label{Phimodeseom}
\end{equation}
where ${\omega^2_k (t)}\equiv \left[ |\vec k|^2 + m_\Phi^2 e^{-F_0\cos(m_F t)}
\right]$.

We shall track the number $N_{\vec{k}}(t)$ of quanta of the {\it initial\/}
state defined via
\begin{equation}
\dot\Phi_{\vec k}(0) = - i\omega_k (0) \Phi_{\vec k}(0).
\label{initialvacuum}
\end{equation}
This defines a vacuum state $|0\rangle$ which is annihilated by the operators
$a_{\vec k}$.

\begin{figure}[!tbp]
\includegraphics{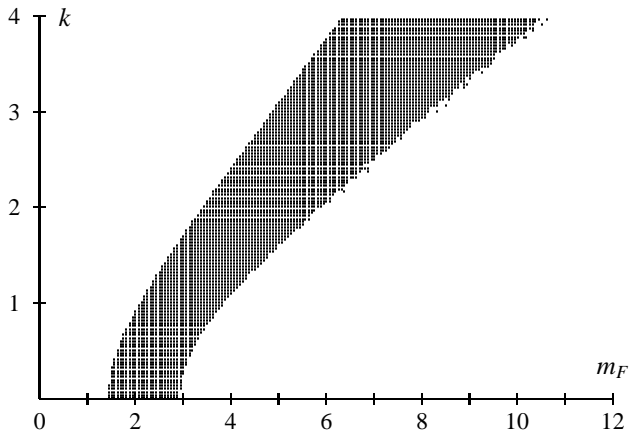}
\caption{The broad instability band for the differential equation
Eq.~(\ref{Phimodeseom}) when $F_0=0.5$.  The dark region shows where the
amplitude of $\Phi_{\vec k}(t)$ grows exponentially.  Both $m_F$ and $k=|\vec
k|$ are expressed in units of $m_\Phi$.\label{bands}}
\end{figure}

The periodicity of the coefficient functions in Eq.~(\ref{Phimodeseom}) under
$t \rightarrow t + 2\pi m_F^{-1}$ implies, by Floquet theory (see
e.g.~\cite{bender}) that there will be values of the dimensionless momentum
$|\vec k|/m_\Phi$ for which $\Phi_{\vec k}(t)$ will undergo exponential
growth, i.e.~there will be unstable bands.  We display this in
Fig.~\ref{bands}.

\begin{figure}[!tbp]
\includegraphics{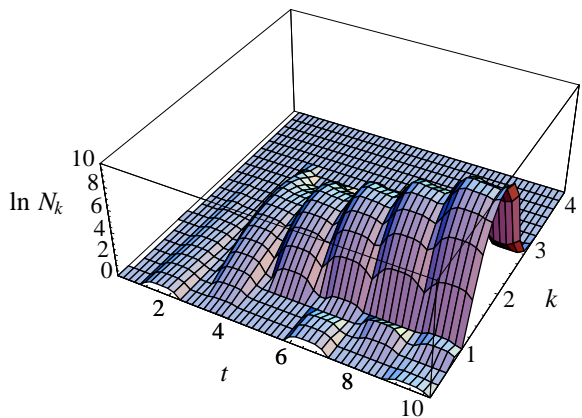}
\caption{$\ln N_{\vec k}(t)$ for $m_F=4m_\Phi$ and $F_0=0.5$.  Note that for
momenta within the instability band of Fig.~\ref{bands}, the number of
particles produced grows exponentially; $t$ and $k$ are expressed in units of
$m_\Phi^{-1}$ and $m_\Phi$, respectively.\label{exponumber}}
\end{figure}

For momenta in the unstable bands, the particle number grows exponentially, as
can be seen from Fig.~\ref{exponumber}, where we plot the expression
\cite{later} for $N_{\vec k}(t)$,
\begin{eqnarray}
&& N_{\vec k}(t) = \nonumber \\
&&\quad
{e^{-F_0}\over 2\omega_k(0) } \left[
e^{2F_0\left[ 1-\cos(m_F t)\right]} |\dot\Phi_{\vec k}(t)|^2
+ \omega_k^2(0) |\Phi_{\vec k}(t)|^2 \right] \ \ \nonumber \\
&&\quad
- {ie^{-F_0\cos(m_F t)}\over 2} \left[ \Phi_{\vec k}^*(t)
\dot\Phi_{\vec k}(t) - \dot\Phi_{\vec k}^*(t) \Phi_{\vec k}(t)
\right] . \ \label{Nfork}
\end{eqnarray}

\begin{figure}[!tbp]
\includegraphics{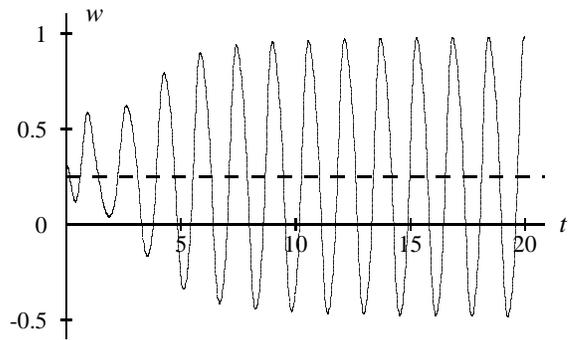}
\caption{The equation of state $w\equiv p/\rho$ as a function of time.  The
dotted line indicates average value of $w$ at later times.  For this case,
$m_F=4m_\Phi$, this average is approximately $0.25$.  $t$ is given in units of
$m_\Phi^{-1}$.
\label{eqofstate}}
\end{figure}

This explosive particle production cannot help but affect brane physics.  We
shall explore the cosmological effects of such particle production in detail
in \cite{later}, but we can make some observations at this point.  First,
these are standard model particles presumably being produced starting when the
brane temperature is ${\cal O}({\rm TeV})$.  While they begin with a
non-thermal equation of state, we should expect that by the time the universe
cools down to nucleosynthesis temperatures, they will have thermalized. 
However, depending on the abundance of these particles, as well as their
equation of state, they can modify the expansion law of the IR brane, i.e.~our
universe.  To follow this effect carefully would require computing the
expectation value of the stress tensor of the produced particles and using it
to source the relevant FRW-like equations that describe the expansion of the
brane.  We have estimated the density and pressure of the produced particles
and plotted the behavior of the ``equation of state'' $w\equiv p/\rho$ as a
function of time in Fig.~\ref{eqofstate}.

The initial radion amplitude will fix the location of the instability band as
will the ratio $m_F/m_\Phi$.  This will change the (time averaged) equation of
state accordingly and thus the expansion rate.

The initial distribution of the Standard Model fields on the brane depends on
the details of the mechanism that produces the IR brane.  If the Standard
Model fields are essentially in their vacuum state, then the subsequent
evolution of the non-thermal particles could differ substantially from the
standard picture of a radiation dominated universe with a temperature
$\alt{\cal O}({\rm TeV})$.  Even with an initial thermal population of fields,
the parametric amplification due to the radion could still produce a
significant fraction of the total energy density in a non-equilibrium
distribution.  Such a large non-equilibrium component at electroweak energies
could produce interesting effects.

We have seen that radion oscillations, corresponding to a ``breather mode''
for the two branes, can give rise to explosive particle production on the IR
brane.  Presumably, similar effects could arise on the UV brane as well,
although the effective field theory techniques used here may not apply. 
Preheating on the UV brane may have even {\it greater\/} consequences for
cosmology on the IR brane, since energy densities on the UV brane can have an
exponentially large effect on IR brane cosmology \cite{branecosmo}.  Radion
induced preheating could also be important in electroweak physics, and in
particular in electroweak baryogenesis \cite{troddenelectroweak}, where this
particle production could affect sphaeleron physics.  We leave all this to our
larger work \cite{later}.

\begin{acknowledgments}
We thank Ira Rothstein for valuable comments.  This work was supported in part
by DOE grant DE-FG03-91-ER40682.
\end{acknowledgments}

\end{document}